# Problems of the matter increase in our observed expansive and isotropic relativistic Universe

**Vladimír Skalský**

Faculty of Materials Science and Technology of the Slovak Technical University, 917 24 Trnava, Slovakia, skalsky@mtf.stuba.sk

**Abstract.** The increase in mass of our observed expansive and isotropic relativistic Universe in the present relativistic cosmology is explained by the extensive assumption of the matter objects emerging on the horizon (of the most remote visibility). However, the physical analysis unambiguously shows that this assumption contradicts the theory of relativity, quantum mechanics, classical mechanics and observations.

### 1. Introduction

In 1929 E. P. Hubble discovered *the expansion of Universe* (Hubble 1929). With expansion of the Universe a set of theoretical problems of *the relativistic cosmology* is connected. One of them is *the problem of matter increase* in *the expansive and isotropic relativistic Universe.*

According to *the Planck quantum hypothesis* (Planck 1899), it has a sense to think on physical parameters of our observed expansive and isotropic relativistic Universe from the moment when it obtained the parameters, which correspond to *the Planck quantities*:

*the Planck mass*

$$m_{Planck} = \sqrt{\frac{\hbar c}{G}} \sim 10^{-8} \text{ kg,} \tag{1}$$

*the Planck length*

$$l_{Planck} = ct_{Planck} = \sqrt{\frac{\hbar G}{c^3}} \sim 10^{-35} \text{ m,} \tag{2}$$

*the Planck time*

$$t_{Planck} = \frac{l_{Planck}}{c} = \sqrt{\frac{\hbar G}{c^5}} \sim 10^{-43} \text{ s.} \tag{3}$$

From the relations (1) and (2) it results

*the Planck mass density*

$$\rho_{Planck} = \frac{c^5}{\hbar G^2} \sim 10^{97} \text{ kg m}^{-3}. \tag{4}$$

At present time these *present parameters of the Universe* are estimated:

*the present mass (of "the visible Universe")*

$$m_{pres} \sim 10^{53} \text{ kg,} \tag{5}$$

*the present gauge factor*

$$a_{pres} \sim 10^{26} \text{ m,} \tag{6}$$

*the present cosmological time*

$$t_{pres} \sim 10^{17} \text{ s} \sim 10^{10} \text{ yr.} \tag{7}$$

From the relations (5) and (6) it results

*the present mass density*

$$\rho_{pres} \sim 10^{-26} \text{ kg m}^{-3}. \tag{8}$$

When comparing the relations (1), (2), (3), (5), (6) and (7) it results that in our expansive ("visible") Universe there grow not only its global space-time dimensions, represented by the gauge factor $a$ and the cosmological time $t$ but its global mass $m$ grows, too.

From comparison of the relations (1) and (5); (2) and (6); and (3) and (7) it results that the Universe mass $m$, the Universe gauge factor $a$ and the Universe cosmological time $t$ (expressed in the Newtonian approximation) during expansive evolution of our Universe increased approximately by 60 ranges.

From comparison of the relations (1) and (5) it results that the global mass of the Universe $m$ increased from approximately one hundred thousandth of gram in the Planck time $t_{Planck}$ (3) to approximately $10^{50}$ tons at the present cosmological time $t_{pres}$ (7).

The increase in mass of the ("visible") Universe $m$ is mathematically given by the fact that – according to the relations (1), (2), (3), (4), (5), (6), (7) and (8) – its volume $V$ increases proportionally to $a^3$, while its mass density $\rho$ decreases proportionally to $a^{-2}$ (and not proportionally to $a^{-3}$).

From comparison of the relation (1), (2), (3), (5), (6) and (7) it results the relations for the mass of the Universe $m$, the gauge factor of the Universe $a$ and the cosmological time of the Universe $t$:

$$m \sim Ca \sim Dt, \tag{9}$$

where $C$ and $D$ are the constants.

From comparison of the relation (2), (3), (6) and (7) it results the relation for the gauge factor of Universe $a$ and the cosmological time of Universe $t$:

$$a \sim ct. \tag{10}$$

From comparison of the relation (9) and the relation (10) it results the relation between constants $C$ and $D$:

$$C \sim \frac{D}{c}. \tag{11}$$

## 2. The Friedmannian models of universe

The isotropic relativistic universe/universes can be described by *the Friedmann general equations of isotropic and homogeneous universe dynamics* (Friedmann 1922, 1924). Using *the Robertson-Walker metrics of isotropic and homogeneous universe* (Robertson 1935, 1936a, b; Walker 1936) they can be expressed as:

$$\dot{a}^2 = \frac{8\pi G a^2 \rho}{3} - kc^2 + \frac{\Lambda a^2 c^2}{3}, \tag{12a}$$

$$2a\ddot{a} + \dot{a}^2 = -\frac{8\pi G a^2 p}{c^2} - kc^2 + \Lambda a^2 c^2, \tag{12b}$$

$$p = K\varepsilon, \tag{12c}$$

where $a$ is the gauge factor, $\rho$ is the mass density, $k$ is the curvature index, $\Lambda$ is the cosmological member, $p$ is the pressure, $K$ is the coefficient of state equation and $\varepsilon$ is the energy density.

The Friedmannian equations (12a), (12b) and (12c) – without introducing of any restrictive supplementary assumptions – have infinite number of solutions, i.e. they describe infinite number of *the Friedmannian models of universe* in the Newtonian (classical-mechanical) approximation.

### 2.1. The standard model of universe

Though the Friedmannian equations (12a), (12b) and (12c) describe infinite number of Friedmannian models of isotropic and homogeneous universe, in the cosmological literature there is a more detailed description only of variants of *the standard model of universe* which are represented by three groups of the Friedmannian models of universe:

1. *The models of universe with predominant dust surroundings*, determined by the Friedmannian equations (12a) and (12b) with $k = +1, 0, -1$ and $\Lambda = 0$ and the state equation:

$$p = 0. \tag{13}$$

2. *The models of universe with predominant radiation (ultra-relativistic particles)*, determined by the Friedmannian equations (12a) and (12b) with $k = +1, 0, -1$ and $\Lambda = 0$ and the state equation:

$$p = \frac{1}{3}\varepsilon. \tag{14}$$

3. *The models of universe with boundary hard-state surroundings*, in which the sound is propagated at *the boundary velocity of signal propagation c*, determined by the Friedmannian equations (12a) and (12b) with $k = +1, 0, -1$ and $\Lambda = 0$ and the state equation:

$$p = \varepsilon. \qquad (15)$$

The parameters of Friedmannian models of universe, which represent the variants of standard model of universe, expressed in the dimensionless conform time $\eta$, defined by the relation:

$$\eta = \pm c \int \frac{dt}{a(t)}, \qquad (16)$$

or

$$c\, dt = a\, d\eta, \qquad (16x)$$

are shown in Tables 1, 2 and 3.

**Table 1.** Parameters of Friedmannian models of the universes with dust state equation $p = 0$ [$0 < \eta < 2\pi$] with $k = +1$; [$0 < \eta < \infty$) with $k = 0, -1$.

| Index of curvature $k$ | +1 | 0 | −1 |
|---|---|---|---|
| Cosmological time $t$ | $\dfrac{a_0}{2c}(\eta - \sin\eta)$ | $\dfrac{a_0}{12c}\eta^3$ | $\dfrac{a_0}{2c}(\sinh\eta - \eta)$ |
| Gauge factor $a$ | $\dfrac{a_0}{2}(1 - \cos\eta)$ | $\dfrac{a_0}{4}\eta^2 = \sqrt[3]{\dfrac{9a_0 c^2}{4}}\sqrt[3]{t^2}$ | $\dfrac{a_0}{2}(\cosh\eta - 1)$ |
| Hubble parameter $H$ | $\dfrac{2c}{a_0}\dfrac{\sin\eta}{(1-\cos\eta)^2}$ | $\dfrac{8c}{a_0}\dfrac{1}{\eta^3} = \dfrac{2}{3}\dfrac{1}{t}$ | $\dfrac{2c}{a_0}\dfrac{\sinh\eta}{(\cosh\eta - 1)^2}$ |
| Mass density $\rho$ | $\dfrac{3c^2}{\pi G a_0^2}\dfrac{1}{(1-\cos\eta)^3}$ | $\dfrac{24c^2}{\pi G a_0^2}\dfrac{1}{\eta^6} = \dfrac{1}{6\pi G}\dfrac{1}{t^2}$ | $\dfrac{3c^2}{\pi G a_0^2}\dfrac{1}{(\cosh\eta - 1)^3}$ |
| Dimensionless density $\Omega$ | $\dfrac{2}{1 + \cos\eta}$ | 1 | $\dfrac{2}{1 + \cosh\eta}$ |

According to Monin *et al*. (1989, p. 108).

**Table 2.** Parameters of Friedmannian models of the universes with ultra-relativistic state equation $p = \dfrac{1}{3}\varepsilon$ [$0 < \eta < \pi$] with $k = +1$; [$0 < \eta < \infty$) with $k = 0, -1$.

| Index of curvature $k$ | +1 | 0 | −1 |
|---|---|---|---|
| Cosmological time $t$ | $\dfrac{a_0}{c}(1 - \cos\eta)$ | $\dfrac{a_0}{2c}\eta^2$ | $\dfrac{a_0}{c}(\cosh\eta - 1)$ |
| Gauge factor $a$ | $a_0 \sin\eta$ | $a_0 \eta = \sqrt{2ca_0}\sqrt{t}$ | $a_0 \sinh\eta$ |
| Hubble parameter $H$ | $\dfrac{c}{a_0}\dfrac{\cotg\eta}{\sin\eta}$ | $\dfrac{c}{a_0}\dfrac{1}{\eta^2} = \dfrac{1}{2}\dfrac{1}{t}$ | $\dfrac{c}{a_0}\dfrac{\cotgh\eta}{\sinh\eta}$ |
| Mass density $\rho$ | $\dfrac{3c^2}{8\pi G a_0^2}\dfrac{1}{\sin^4\eta}$ | $\dfrac{3c^2}{8\pi G a_0^2}\dfrac{1}{\eta^4} = \dfrac{3}{32\pi G}\dfrac{1}{t^2}$ | $\dfrac{3c^2}{8\pi G a_0^2}\dfrac{1}{\sinh^4\eta}$ |
| Dimensionless density $\Omega$ | $\dfrac{1}{\cos^2\eta}$ | 1 | $\dfrac{1}{\cosh^2\eta}$ |

According to Monin *et al*. (1989, p. 109).

**Table 3.** Parameters of Friedmannian models of the universes with boundary hard-state equation $p = \varepsilon$ $[0 < \eta < \pi/2]$ with $k = +1$; $[0 < \eta < \infty)$ with $k = 0, -1$.

| Index of curvature $k$ | +1 | 0 | −1 |
|---|---|---|---|
| Cosmological time $t$ | $\dfrac{a_0}{c}\displaystyle\int_0^\eta \sqrt{\sin 2\xi}\,d\xi$ | $\dfrac{2\sqrt{2}a_0}{3c}\eta\sqrt{\eta}$ | $\dfrac{a_0}{c}\displaystyle\int_0^\eta \sqrt{\sinh 2\xi}\,d\xi$ |
| Gauge factor $a$ | $a_0\sqrt{\sin 2\eta}$ | $a_0\sqrt{2\eta} = \sqrt[3]{3ca_0^2}\sqrt[3]{t}$ | $a_0\sqrt{\sinh 2\eta}$ |
| Hubble parameter $H$ | $\dfrac{c}{a_0}\dfrac{\cot g\, 2\eta}{\sqrt{\sin 2\eta}}$ | $\dfrac{c}{2\sqrt{2}a_0}\dfrac{1}{\eta\sqrt{\eta}} = \dfrac{1}{3}\dfrac{1}{t}$ | $\dfrac{c}{a_0}\dfrac{\cot gh\, 2\eta}{\sqrt{\sinh 2\eta}}$ |
| Mass density $\rho$ | $\dfrac{3c^2}{8\pi G a_0^2}\dfrac{1}{\sin^3 2\eta}$ | $\dfrac{3c^2}{64\pi G a_0^2}\dfrac{1}{\eta^3} = \dfrac{1}{24\pi G}\dfrac{1}{t^2}$ | $\dfrac{3c^2}{8\pi G a_0^2}\dfrac{1}{\sinh^3 2\eta}$ |
| Dimensionless density $\Omega$ | $\dfrac{1}{\cos^2 2\eta}$ | 1 | $\dfrac{1}{\cosh^2 2\eta}$ |

According to Monin *et al.* (1989, p. 109).

## 2.2. The model of expansive non-decelerative universe

Only Friedmannian model of universe with the Newtonian non-modified relations is *the (flat) expansive nondecelerative (isotropic and homogeneous) universe* (ENU) (Skalský 1992, 1991, 1997, 1998, 2000d, e).

The model of ENU can be determined in four ways (Skalský 1997). One of these ways of determination of the model of ENU is based on a possibility of unambiguous determination of the beginning conditions of the expansive and isotropic relativistic–quantum-mechanical universe (Skalský, 1997, 2000b, c, d).

In 1970 S. W. Hawking and R. Penrose proved that the expansive universe had to begin its expansive evolution by the singularity (Hawking and Penrose 1970). It means that the expansive and isotropic relativistic universe, which began its expansive evolution by the singularity – according to *the general theory of relativity* – could begin its expansive evolution at only possible velocity: at the boundary velocity of signal propagation.

The matter-space-time in the expansive and isotropic relativistic universe, which began its expansive evolution from the cosmological singularity was quantified, therefore, the hypothetical assumption on the beginning velocity of relativistic universe expansion must corresponds with the velocity, which is permitted *the quantum theory* in the Planck time $t_{Planck}$ (3).

According to the relations, which determine the Planck length $l_{Planck}$ (2) and the Planck time $t_{Planck}$ (3), the universe can begin its expansive evolution at only possible velocity: at the boundary velocity of signal propagation $c$ (Skalský 2000b, c, e).

In the expansive and isotropic relativistic universe for the space-time relations are valid

*the Lorentz transformations*:

$$x' = \frac{x - vt}{\sqrt{1 - \dfrac{v^2}{c^2}}}, \tag{17}$$

$$y' = y, \tag{18}$$

$$z' = z, \tag{19}$$

$$t' = \frac{t - \dfrac{v}{c^2}x}{\sqrt{1 - \dfrac{v^2}{c^2}}}, \tag{20}$$

where $x'$, $y'$, $z'$ are the space co-ordinates and $t'$ is the time in the co-ordinate system $\Sigma'$; $x$, $y$, $z$ are the space co-ordinates and $t$ is the time in another inertial co-ordinate system $\Sigma$; and $v$ is the velocity of the system $\Sigma'$ with respect to the system $\Sigma$.

In *the Newton theory of gravitation (the classical mechanics)* for the space-time relations are valid *the Galileian transformations*:

$$x' = x - vt, \quad (21)$$

$$y' = y, \quad (22)$$

$$z' = z, \quad (23)$$

$$t' = t. \quad (24)$$

From the Lorentz transformation (20) it results a relativistic dilatation of the time-flow depending on the velocity of moving objects, and at the boundary velocity of signal propagation $c$ the time-flow will stop.

However, according to the Galileian transformation (24), in the Newtonian classical mechanics the time flows equally in all co-ordinate systems without regard to their mutual velocity, or any next assumptions.

From these facts it results unambiguously that:

**In the expansive and isotropic relativistic universe, which began its expansive evolution at the boundary velocity of signal propagation (at which the time-flow will stop), the gauge factor *a*, expressed in the Newtonian approximation, must grow at the constant velocity $v = c$, during the whole expansive evolution** (Skalský 1997, 1999e).

It means that for the gauge factor $a$ and the cosmological time $t$ of the expansive and isotropic relativistic universe, which began its expansive evolution at the boundary velocity of signal propagation (at which the time-flow will stop), in the Newtonian approximation is valid the relation (Skalský 1992, 1991, 2000d):

$$a = ct. \quad (25)$$

The Friedmannian equations (12a), (12b) and (12c) with the value of index curvature $k = 0$, the value of cosmological member $\Lambda = 0$ and the restrictive condition determined by the relation (25) have solution only with the value of state equation coefficient $K = -1/3$. It means that the Friedmannian model of the expansive non-decelerative (isotropic and homogeneous) universe (ENU) is determined unambiguously by the Friedmannian equation (12a) and (12b) with $k = 0$ and $\Lambda = 0$ and

*the zero gravitational force state equation* (Skalský 1991, 2000d):

$$p = -\frac{1}{3}\varepsilon. \quad (26)$$

The parameters of Friedmannian model of the ENU, expressed in the dimensionless conform time $\eta$, defined by the relation (16), or (16x), are shown in Table 4.

**Table 4.** Parameters of Friedmannian model of the ENU with state equation $p = -\frac{1}{3}\varepsilon$ [$0 < \eta < \infty$).

| | |
|---|---|
| Index of curvature $k$ | 0 |
| Cosmological time $t$ | $t_0 e^\eta = \dfrac{a}{c}$ |
| Gauge factor $a$ | $ct_0 e^\eta = ct$ |
| Hubble parameter $H$ | $\dfrac{e^{-\eta}}{t_0} = \dfrac{1}{t}$ |
| Mass density $\rho$ | $\dfrac{3e^{-2\eta}}{8\pi G t_0^2} = \dfrac{3}{8\pi G t^2}$ |
| Dimensionless density $\Omega$ | 1 |

According to Skalský (1991, p. 318).

The parameters of the ENU – determined by the Friedmannian equations (12a) and (12b) with $k = 0$ and $\Lambda = 0$ and the state equation (26) – are mutually linearly linked, therefore they can be expressed without problems in mutually relations. For transparency we express the parameters of the ENU in all-possible relations and variants (Skalský 1997, 2000d, e):

$$a = ct = \frac{c}{H} = \frac{2Gm}{c^2} = \sqrt{\frac{3c^2}{8\pi G\rho}}, \tag{27}$$

$$t = \frac{a}{c} = \frac{1}{H} = \frac{2Gm}{c^3} = \sqrt{\frac{3}{8\pi G\rho}}, \tag{28}$$

$$H = \frac{c}{a} = \frac{1}{t} = \frac{c^3}{2Gm} = \sqrt{\frac{8\pi G\rho}{3}}, \tag{29}$$

$$m = \frac{c^2 a}{2G} = \frac{c^3 t}{2G} = \frac{c^3}{2GH} = \sqrt{\frac{3c^6}{32\pi G^3 \rho}}, \tag{30}$$

$$\rho = \frac{3c^2}{8\pi G a^2} = \frac{3}{8\pi G t^2} = \frac{3H^2}{8\pi G} = \frac{3c^6}{32\pi G^3 m^2} = -\frac{3p}{c^2}, \tag{31}$$

$$p = -\frac{c^4}{8\pi G a^2} = -\frac{c^2}{8\pi G t^2} = -\frac{c^2 H^2}{8\pi G} = -\frac{c^8}{32\pi G^3 m^2} = -\frac{c^2 \rho}{3} = -\frac{1}{3}\varepsilon. \tag{32}$$

where $t$ is the cosmological time, $H$ is the Hubble coefficient ("constant"), and $m$ is the mass of the ENU.

From the relations (30) it results the relations for the ENU mass $m$, the ENU gauge factor $a$ and the ENU cosmological time $t$:

$$m = Ca = Dt, \tag{33}$$

where $C$ and $D$ are the (total) constants:

$$C = \frac{m}{a} = \frac{m}{ct} = \frac{Hm}{c} = \sqrt{\frac{8\pi G\rho m^2}{3c^2}} = \frac{c^2}{2G} = 6.734\,67(15) \times 10^{26}\ \mathrm{kg\,m^{-1}}, \tag{34}$$

$$D = \frac{cm}{a} = \frac{m}{t} = Hm = \sqrt{\frac{8\pi G\rho m^2}{3}} = \frac{c^3}{2G} = 2.019\,00(37) \times 10^{35}\ \mathrm{kg\,s^{-1}}, \tag{35}$$

where

$$C = \frac{D}{c}. \tag{36}$$

### 3. The problem of the Universe mass increase

In all models of the expansive and isotropic relativistic Universe the matter (the matter objects) do not exist eternally, but originated (was created) in certain phase of its expansive evolution.

In the models of expansive Universe from the pre-inflationary period of the relativistic cosmology, and in which the Planck quantum hypothesis is respected, the origin (creation) of matter is considered approximately at the Planck time $t_{Planck}$ (3).

In the models of expansive Universe with the inflationary evolution phase the matter originated (was created) in later period. In *the Linde model of chaotic inflationary Universe* (i.e. in the only model of inflationary models of the universe which is considered as viable) the matter originated (was created) at the end of inflationary evolution phase, i.e. approximately at the cosmological time $t \sim 10^{-37}$ s (Linde 1990).

In the model of ENU the matter originates (is created) during the whole expansive evolution. Like in *the Bondi-Gold-Hoyle steady-state theory* (Bondi and Gold 1948; Hoyle 1948), but with that essential difference that while Bondi, Gold and Hoyle introduced the creation of matter on the basis of a hypothetical assumption that the observed expansive and isotropic relativistic Universe has a constant (immutable) mass density $\rho$ – however, this assumption contradicts the present observations – *the permanent constant maximum possible creation of matter C* (34), or *D* (35), resulted from the model properties of ENU; namely from the deterministic linear mutual dependence of the mass $m$, the gauge factor $a$, the cosmological time $t$ (and other parameters of ENU) (Skalský 1997, 2000d).

**The relation (1), (2), (3), (4), (5), (6), (7) and (8), which are presented in the cosmological literature, are not derived on the basis of any variants of the standard model of universe!**

It is almost generally known (indeed it results from designation of beginning parameters of Universe) that the parameters (1), (2), (3) and (4) are derived on the basis of the Planck quantum hypothesis, using the dimensional analysis and the constants $G$, $c$ and $\hbar$, or $h$.

But, it is less known that the present parameters of Universe (5), (6), (7) and (8) are derived on the basis of the classical mechanics (the Newton theory of gravitation) for a hypothetical Newtonian flat homogeneous Universe, determined by the Newtonian (non-modified) relation for the escape velocity

$$v_{escape} = \sqrt{\frac{2Gm}{r}} \qquad (37)$$

with corresponding mutually linearly linked parameters, i.e. they are determined for the mass of homogeneous matter sphere $m$ with the radius $r$ (where $r = a$) and with the escape velocity

$$v_{escape} \sim c \qquad (38)$$

from its surface. As a known quantity we can put the value of any of the parameters of Universe: $m$, $a$, $t$, or $H$ (where $H = 1/t$), $\rho$, or $\varepsilon$ (where $\varepsilon = \rho c^2$). The value of other quantities (parameters) will result from the above.

Only the Friedmannian model of universe with the Newtonian non-modified relations is the model of ENU (Skalský 1992, 1991, 1993, 1994, 1997, 1998, 2000d, e).

It means that:

**The present parameters of universe (5), (6), (7) and (8) – derived by means of the Newtonian non-modified relations (37) and (38) – are *de facto* derived on the basis of ENU, without knowing this fact!**

As we have shown, according to the standard model of universe without inflationary evolution phase, the matter originated approximately in the Planck time $t_{Planck} \sim 10^{-43}$ s. Therefore, for physical explanation of linear constant increase in mass of observed expanding and isotropic relativistic Universe – which occurs in the time of its whole expansive evolution – into the standard model of universe there was introduced a hypothetical extensive supplementary assumption that the present observed Universe began its expansive evolution from the area no less than 29 ranges larger than the Planck length $l_{Planck}$ (2), i.e. approximately from the area with the dimension $l \sim 10^{-6}$ m.

According to this hypothetical extensive supplementary assumption, the dimensions of "visible" Universe increase at the velocity $v \sim c$, but its matter structure expands more slowly. Therefore, the total mass of "visible" Universe (in the Newtonian approximation) $m$ increases in the process of its expansive evolution. The increase in mass of the "visible" Universe – according to this hypothetical extensive supplementary assumption – is caused by *the extensive emergence of matter objects* on *the (Newtonian-Euclidean) horizon (of the most remote visibility)*.

According to *the special theory of relativity* (Einstein 1905), the velocity of signal propagation is limited by the boundary velocity of signal propagation $c$. Therefore, by introduction of the hypothetical extensive supplementary assumption into the standard model of universe, i.e. assumption that the mass of expansive Universe grows as a result of emergence of matter objects on the Newtonian-Euclidean horizon (of the most remote visibility), the so-called *problem of horizon (causality) of Universe*, arose in the standard model of universe.

The problem of Universe horizon may be formulated by the next question: If the observed Universe began its expansive evolution from the area with the dimension $l \sim 10^{-6}$ m, then the present "visible" Universe must represent $(10^{29})^3 = 10^{87}$ causally not bounded areas in the Planck time $t_{Planck}$ (3). However, how is it possible that the evolution of our expansive Universe began simultaneously in such a great number of causally non-bounded, but mutually evolutionarily very precisely co-ordinated areas?

In the standard model of universe from the hypothetical extensive supplementary assumption on the increase in mass of expansive Universe as a result of emergence of mass objects on the hypothetical Newtonian-Euclidean horizon of the most remote visibility it results that the present mass density of our Universe $\rho_{pres}$ is dependent on its beginning mass density $\rho_{beg}$.

Under this hypothetical extensive assumption the retrospective extrapolation of expansive evolution of Universe leads to the unambiguous conclusion that in the Planck time $t_{Planck}$ (3) our expansive Universe must had to have the beginning mass density $\rho_{beg} = (1 \pm 10^{-59})\rho_c$, where $\rho_c$ is a critical mass density. Because, if $\rho_{beg}$ had been larger, its expansive evolution would have been replaced by the contraction. If $\rho_{beg}$ had been smaller, no hierarchic gravitationally-bounded rotational systems (HGRS) in the Universe would have arise, i.e. no life would have arise in the Universe, and so no mankind would exist today.

The question, why the beginning mass density $\rho_{beg}$ in our expansive and isotropic Universe was – under above shown hypothetical extensive assumption – critical with such high precision, is usually designated in the cosmological literature as *the problem of flatness (Euclidicity) of Universe*.

From the definition of the Planck length $l_{Planck}$ (2), the Planck time $t_{Planck}$ (3) and the problem of Universe flatness it results that in the Planck time $t_{Planck}$ (3) the matter, concentrated in each of the above mentioned approximately $10^{87}$ originally independent areas ("the Plancktons"), had to expand at

*the Planck velocity*

$$v_{Planck} = c \qquad (39)$$

at the distance $r = l_{Planck}$ (2).

In the cosmological literature the concept *horizon* is used in two fundamental meanings, which must be distinguished. We must distinguish between:

1. *the optical horizon (the horizon of visibility, the horizon of particles),*

2. *the horizon of (all) events (the horizon of the most remote visibility).*

These concepts are often used without distinction. For example, the horizon is often simply referred to without closer determination (similarly as it has been – intentionally – stated above, in the cosmological literature the term *the problem of horizon of Universe* is mostly used). Therefore, it comes to confusions about their fundamental meanings. Besides these confusions it also comes to confusions about the Newtonian-Euclidean meanings (statements) with their relativistic equivalents.

To prevent misunderstanding, before we start our own analysis of the problem of mass increase in observed Universe we characterise (define) the meanings of individual concepts.

The linearly mutually bounded relations (9), (10) and (11) which result from parameters (1), (2), (3), (5), (6) and (7) are very near to the linearly mutually bounded relations (27), (28), (29), (30), (31) and (32), or (33), derived on the basis of Friedmannian model of the ENU. They are distinguished only by the mark of equality. Therefore, concrete significance of concepts 'the optical horizon', 'the horizon of events' (and any next with them bounded concepts) will be illustrated on the basis of the ENU model (i.e. on the basis of the simplest Friedmannian model and simultaneously the only Friedmannian model of universe with the Newtonian non-modified relations).

The Friedmannian model of the ENU during the whole expansive evolution expands at the boundary velocity of signal propagation *c*. According to the Einstein special theory of relativity (Einstein 1905), the boundary velocity of signal propagation *c* is independent of the moving objects velocity, therefore, each observer in the ENU is in its "centre" (irrespectively of the velocity of its movement relative to other cosmological objects and relative to the cosmic background, represented by the cosmic background radiation). In the Newtonian approximation he can imagine it as an expanding sphere. The observer is in its centre and the remotest objects on the surface of this sphere expand from him at the velocity

$$v = c \ . \qquad (40)$$

*The gauge (scale) factor a* in the model of ENU is the distance of the observer from the expanding sphere surface, i.e. in the model of ENU is valid the relation:

$$a = r \ , \qquad (41)$$

where *r* is the radius of homogeneous matter sphere, expanding at the velocity *v* (40).

In the flat expanding homogeneous Newtonian universe with the Newtonian non-modified relations (i.e. in the model of ENU) the velocity *v* (40) is the exact escape velocity, therefore, the relation (41) can be extended on *the Schwarzschild gravitational radius* $r_g$:

$$a = r = r_g \ . \qquad (42)$$

The light propagates at the finite velocity in the Newton theory of gravitation (the classical mechanics), too. (The finite velocity of light was discovered by O. Ch. Römer in 1676, i.e. 11 years before publication of the famous I. Newton work *Philosophiae naturalis principia mathematica* (Newton 1687).) It means that the gauge factor *a* in the Newtonian model of the flat expansive homogeneous universe (with the Newtonian non-modified relations) (i.e. in the ENU) simultaneously represents also *the horizon of visibility (the optical horizon)* $h_v$. Therefore, in the case of the Friedmannian model of the ENU the equations (42) can be further extended on the next relations:

$$a = r = r_g = h_v \ . \qquad (43)$$

The optical horizon $h_v$ in the ENU represents a connecting line (from point of view of optical signal propagation) of relatively simultaneous events. (For example the events which the observer on the Earth optically observes as simultaneous. But based on his astronomical knowledge he knows that what he is now optically observing as simultaneous, was realised for example on the Moon more than 1 second ago, on the Sun more than 8 minutes ago, on the nearest star Proxima Centauri more than 4 years ago … etc.)

In Figure 1 we pictured the evolution of the ENU in the form of the time cone (Skalský 1991, 1994, 1997, 2000e).

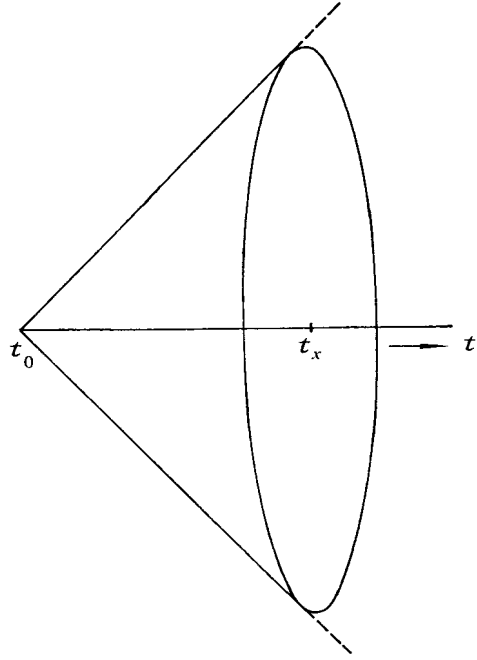

**Figure 1.** The 3-dimensional space-time projection of the 4-dimensional expansive and isotropic relativistic universe in the Newtonian approximation at any cosmological time $t_x$.

According to the Euclidean geometrical point of view, the mass in the model of Newtonian-Euclidean flat expansive homogeneous universe might grow in three ways:

1. *Extensive*. As the results of emergence of matter objects on the horizon of events, i.e. on the surface of the 3-dimensional expanding Newtonian-Euclidean homogeneous matter sphere, which in the 2-dimensional projection in Figure 1 corresponds the ellipse.

2. *Intensive*. As the results of creation of matter objects during the whole expansive evolution of universe, represented by the emergence of matter objects on the whole horizon of visibility $h_v$ ($= a = r = r_g$), which, in full Newtonian-Euclidean expression, corresponds an arbitrary connecting line (abscissa), connecting the point $t_x$ (in which is the observer) with an arbitrary point on the surface of the expanding sphere. In the 3-dimensional projection in Figure 1 to the above abscissa corresponds the connecting line of point $t_x$ with an arbitrary point on the ellipse.

3. *Extensive-intensive*. Simultaneous application of both the extensive and intensive mechanisms, i.e. one part of matter objects emerges on the horizon of events and another part of matter objects emerges on the horizon of visibility (as a result of the permanent creation of matter).

In Figure 2 the 2-dimensional special-relativistic (half) projection of the 4-dimensional expansive and isotropic relativistic universe is projected into the 2-dimensional Newtonian (half) projection of the expansive and isotropic relativistic universe (Skalský 1993, 1994, 1997, 2000e). This projection enables us to very plastically compare the differences of the special-relativistic properties of the universe and their expressions in the Newtonian approximation.

In Figure 2 the ENU gauge factor $a$ ($= r = r_g = h_v$) in the time $t_x$ represents the connecting line (abscissa) between the point $t_x$ (in which is the observer) and the point $A$.

The horizon of events of ENU, representing in the Newtonian approximation the surface of the expansive Euclidean sphere, to which in the 3-dimensional Newtonian projection of ENU in Figure 1 corresponds the ellipse, is in Figure 2 represented by the point $A$.

In the special-relativistic projection of the expansive and isotropic relativistic universe the situation looks substantially differently as than in its Newtonian projection. In Figure 2 to the (Newtonian) gauge factor $a$ ($= r = r_g = h_v$), representing the abscissa bordered by points $t_x$ and $A$, in the special-relativistic projection of the expansive and isotropic relativistic universe, corresponds the curve which connects the point in location of observer $t_x$ with the point $A'$ ($\equiv t_0$) which connects the special-relativistic relatively simultaneous events.

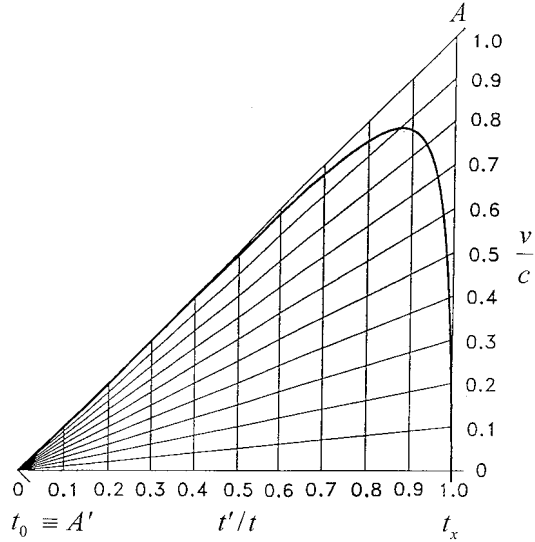

**Figure 2.** The 2-dimensional Newtonian (half) space-time projection and simultaneously the 2-dimensional special-relativistic (half) space-time projection of the 4-dimensional expansive and isotropic relativistic universe at any cosmological time $t_x$.

In the expansive and isotropic relativistic universe it comes not only to the special-relativistic contraction of radial length of individual expanding objects, but – as a result of the special-relativistic dilatation of time – also to the special-relativistic contraction of its total dimensions. The point of intersection of connecting line of point $t_0$ and points $t_x \equiv 0, 0.1, \dots 0.9, 1.0 \equiv A$ with the curve, which represents the special-relativistic dilatation of time in the expansive and isotropic relativistic universe, clearly presents the special-relativistic contraction of the radial dimension of the expansive and isotropic relativistic universe, too.

In Figure 2 we can see very clearly, why the total parameters of the expansive and isotropic relativistic universe have finite non-zero values only in the Newtonian approximation, and we may easily verify the fact why the parameters of the expansive and isotropic relativistic universe must be expressed in the Newtonian approximation.

If we compare the space-time properties of the expansive and isotropic relativistic universe, expressed in a special-relativistic way and in the Newtonian approximation, we can see that in smaller distances they differ only insignificantly, but in larger distances the difference steeply grows. The expansive and isotropic special-relativistic universe projected into the Newtonian projection of expansive and isotropic relativistic universe has the largest dimensions in the distance $r = 78.388\% \, a$ (Skalský 1994, 1997, 2000e). In the (Newtonian) larger distances the special-relativistic dimensions of the expansive and isotropic relativistic universe steeply drop and in the distance of gauge factor $a \, (= r = r_g = h_v)$ its values are zero.

The expansive and isotropic relativistic universe, expanding at the velocity $v$ (40) as a result of the special-relativistic dilatation of time has all total matter-space-time values equal zero, i.e. it is realised singularly (Skalský 1997).

According to the Planck quantum hypothesis (Planck 1899) on the actual physical quantities of universe has the physical sense considered in the Planck time $t_{Planck}$ (3). It means that all its total parameters are equal zero, but its local parameters are non-zero and relative (i.e. they exist only in the scope concrete actual relativistic relations).

With the velocity $v$ (40) – as a result of the Lorentz special-relativistic dilatation of time – the time-flow was stopped, and the relativistic universe is closed in space-time manner. Therefore, the increase in its mass as the result of the emergence of matter objects on the horizon of events principally cannot occur! It means that the increase in mass of the expanding relativistic Universe may be only the result of the permanent creation of matter.

In the present time the singular relativistic–quantum-mechanical closing of expanding and isotropic relativistic universe is ignored. It is argumented just by the increase of its mass. The hypothetical assumption of emergence of matter on the horizon of events was into the relativistic cosmology introduced just from this reason.

In this connection it must be openly stated that:

**The permanent constant maximum possible creation of matter $C$ (34), or $D$ (35), results from the relations (30), or from its variants (27), (28), (29), (31) and (32), or (33), and from the model properties of ENU. The permanent creation of matter $C$ (34), or $D$ (35), was not observed up to this time and no**

**physically acceptable theoretical mechanism/mechanisms was suggested which could explain the creation of matter.**

The fact that the whole of the observed Universe can be described (without limit, i.e. zero values) only in the Newtonian approximation, however, it does not mean that the Universe is Newtonian. The expansive and isotropic relativistic Universe, which began its expansive evolution at the escape velocity, determined by the relation (39), is flat in the Newtonian approximation, but in the special-relativistic description is spherically symmetrically and singularly closed in space-time manner (in the special-relativistic description is flat only asymptotically, when $a \to \infty$). (See Figure 2.) It means that any consideration on the hypothetical Newtonian-Euclidean horizon of events in the expanding and isotropic relativistic universe contradicts the general theory of relativity. Therefore, any consideration on the hypothetical actual neighbouring "Plancktons", or on any hypothetical actual matter objects behind the hypothetical Newtonian-Euclidean horizon of events in the expanding and isotropic relativistic universe are inadmissible (they have no physical sense).

The hypothetical extensive assumption on the emerging of matter object on the hypothetical Newtonian-Euclidean horizon of events is in contradiction to:

1. The axiomatic principles on which the Einstein general theory of relativity is constructed.

2. The relations of the Lorentz special-relativistic transformation.

By these conclusions the analysis of the problems of extensive increase in mass of the standard model of universe could be concluded. However, in view of the fact that the hypothetical extensive assumption on increase in mass of Universe in the result of emergence of matter objects on the hypothetical Newtonian-Euclidean horizon of events constitutes an essential part of *the present cosmological paradigm*, our analysis will be extended to the non-relativistic possibilities.

In view of the fact that the Newton theory of gravitation is the special partial solution of the Einstein theory of gravitation (the general theory of relativity) and the equivalence of the gravitational mass and the inertial mass is valid in it (though it is not specifically stressed), the hypothetical assumption on the emergence of matter objects on the hypothetical Newtonian-Euclidean horizon of events contradicts the Newton theory of gravitation, too. From this fact it results that non-possibility of increase in mass of expansive Universe as a result of emergence of matter objects on the hypothetical Newtonian-Euclidean horizon of the most remote visibility must lead to the contradictions even at the assumption that the expansive and isotropic Universe will be considered as Newtonian. – Id est, if we investigate it in the whole extent of Newtonian admissible parameters, so, as well as behind the limits (scope) of correspondence of the Einstein theory of gravitation (the general theory of relativity) and the Newton theory of gravitation (the classical mechanics).

Therefore, let us admit that the observed Universe is not relativistically closed in space-time manner, but have the Newtonian-Euclidean properties, i.e. that the considerations on emergence of matter objects on the hypothetical Newtonian-Euclidean horizon of events have the physical sense.

From the Newtonian (the classical mechanics) relations for the mass of homogeneous matter sphere $m$ with the radius $r$ and the mass density $\rho$:

$$m = \frac{4}{3}\pi r^3 \rho \qquad (44)$$

it results that the chosen starting mass of universe $m_0$ in the Newtonian-Euclidean expansive homogeneous universe – the mass of which increases as a result of emergence of matter objects on the horizon of the most remote visibility – could be extended to the volume of homogeneous matter sphere with the resulting mass density of universe $\rho_x$ and the resulting radius (Skalský, 1999):

$$r_x = \sqrt[3]{\frac{3m_0}{4\pi\rho_x}} \; . \qquad (45)$$

The matter objects in the Newtonian-Euclidean expansive homogeneous universe – in which the matter objects emerge on the horizon of the most remote visibility – and, according to the relation (10), expand at the velocity $v \sim c$, when they appear in the distance $r_x$, expand at the velocity (Skalský, 1999)

$$v_{r(x)} \sim \frac{r_x}{a_x}c, \qquad (46)$$

where $a_x$ is the corresponding result gauge factor.

From the relations (1), (8) and (45) it results that the mass of "Planckton" $m_{Planck}$ (1) at the present cosmological time $t_{pres}$ (7) could be extended to the volume of homogeneous matter sphere with the present mass density $\rho_{pres}$ (8) and with the radius (Skalský, 1999)

$$r_{Planck-pres} \sim 10^6 \, \text{m}. \qquad (47)$$

From the relations (6), (46) and (47) it results that the velocity of the initial "Planckton" $v = c$ in the homogeneous Universe with the present gauge factor $a_{pres}$ (6) in the distance of the homogeneous matter sphere with the radius $r_{Planck-pres}$ (47) had to decrease to the velocity (Skalský, 1999)

$$v_{r(Planck-pres)} \sim 10^{-12} \text{ ms}^{-1} \sim 10^{-20} c. \tag{48}$$

In the Planck time $t_{Planck}$ (3), the matter-space-time was quantified. Therefore, it means that if the total mass of Newtonian-Euclidean universe increased in the result of emergence of matter objects on the hypothetical Newtonian-Euclidean horizon of events, it would be possible only under assumption that the relative velocity of the neighbouring "Plancktons" was lower than the velocity of expansion of separate "Plancktons", determined by the relation (39).

But, this assumption would, however, mean a catastrophe because in the whole assumed area with the dimension $l \sim 10^{-6}$ m, immediately after the cosmological time $t_{Planck}$ (3), i.e. at the cosmological time $2t_{Planck}$, among neighbouring "Plancktons" it would have to lead to their mutual collision.

In each of "the Plancktons" was the maximum possible of mass density $\rho_{Planck}$, determined by the relation (4). Therefore, the result of rapid slowing down of their expansion, caused by their mutual collisions, would be transformation of the flat expanding homogeneous and isotropic relativistic Universe into the contracting one and next to the (final cosmological) singularity (Skalský 1993, 1997).

If we ignored this fact (for example justifying that the beginning conditions of expansive evolution of Universe are not known; or that the beginning conditions of Universe evolution cannot be determined by the return extrapolation from the present conditions; that to the beginning state of Universe the theory of relativity or the classical mechanics cannot be applied because proceed the quantum-mechanical effects occurred in it; and so on), even so the model properties of the Newtonian universe with the Euclidean geometry and with the relativistic properties of moving matter objects would unambiguously exclude the hypothetical assumption on growing the mass of Universe in the result of emergence matter objects on the hypothetical Newtonian-Euclidean horizon of events.

The matter objects with the relativistic properties, which would emerge on the hypothetical Newtonian-Euclidean horizon of events in this case would have to move exactly at the velocity

$$v = c^*, \tag{49}$$

where $c^*$ is the velocity of light in the matter surroundings.

For the moving special-relativistic mass $m'$ and for the proper (rest) mass $m$ of the matter object in the inertial co-ordinate system, moving relatively to the observer at the velocity $v$ is valid

*the Einstein transformation*:

$$m' = \frac{m}{\sqrt{1 - \frac{v^2}{c^2}}}. \tag{50}$$

According to the Einstein general theory of relativity, the matter objects with relativistic properties, emerging on the hypothetical Newtonian-Euclidean horizon of events and moving at the boundary velocity of signal propagation in the mass surrounding $v = c^*$, would have infinite relativistic mass (energy). It means that the hypothetical Newtonian-Euclidean universe, in which the matter objects with relativistic properties emerge on the hypothetical Newtonian-Euclidean horizon of events principally cannot exist.

Nor on the assumption that we would ignore observations on the relativistic properties of matter objects and we would assume that the matter objects have the Newtonian properties, i.e. that for their mass is valid

*the Newtonian transformation*:

$$m' = m, \tag{51}$$

it would be not possible to explain the increase of universe mass by the hypothetical extensive assumption on the emergence of matter objects on the hypothetical Newtonian-Euclidean horizon of events.

The situation with the mass increase of universe as a consequence of emergence of matter objects on the hypothetical Newtonian-Euclidean horizon of events in the Newtonian universe with the Euclidean geometry, in which the signals propagate at the velocities

$$v \leq c, \tag{52}$$

in which for the mass of moving objects the transformation (51) is valid, and in which we would neglect the gravitational interaction, would be simple, if only the hypothetical Newtonian-Euclidean horizon of events grow and the matter objects emerging on this horizon would not move (i.e. when the distances among the matter objects would not change and, therefore, so nor the mass density of expansive universe $\rho$ would change either).

But, the problem is – besides others – in this, that, according to the relation (10), the matter objects in the Universe at the distance of gauge factor $a$ expand at the velocity $v \sim c$, therefore, the matter objects in the hypothetical Newtonian-Euclidean universe, in which the matter objects emerge on the hypothetical Newtonian-Euclidean horizon of events in the distance of gauge factor $a$ would must expand at the velocity $v \sim c$. Since in the observed Universe the mass density $\rho$ decreases proportionally to $a^{-2}$, the increase in mass in the hypothetical Newtonian universe with the Euclidean geometry and the matter objects, for which is valid the Galileian transformation (51), would be possible to explain by the hypothetical assumption emergence of matter objects on the hypothetical Newtonian-Euclidean horizon of events only on the assumption that the velocity of its expansion would decrease rapidly.

As shown, the velocity of matter objects in the Planck (or more precisely in immediately post-Planck) time expanding at the velocity $v = c$, according to the relation (48), would have to up to the present time decrease to the velocity $v \sim 10^{-20} c$.

Equally – as we determined the relations (47) and (48) – we can determine the relations for any arbitrary dimensions of Universe with the hypothetical Newtonian-Euclidean properties in which the matter objects emerge on the hypothetical Newtonian-Euclidean horizon of events.

For example, at

*the cosmological time of Universe at the end of radiation (photon) era*

$$t_{end} \sim 3 \times 10^5 \text{ yr.} \tag{53}$$

The flat expansive homogeneous Universe at the time $t_{end}$ (53) – according to the relation (10) – had

*the gauge factor of Universe at the end of radiation era*

$$a_{end} \sim 3 \times 10^{21} \text{ m,} \tag{54}$$

and – according to the relations (9) and (35) – had

*the mass of Universe at the end of radiation era*

$$m_{end} \sim 2 \times 10^{48} \text{ kg.} \tag{55}$$

According to the relation (45), the mass of the Universe at the end of radiation era $m_{end}$ (55) at the present cosmological time $t_{pres}$ (7) could be extended to the volume of homogeneous matter sphere with the present mass density $\rho_{pres}$ (8) and with the radius (Skalský 1999)

$$r_{end-pres} \sim 3 \times 10^{24} \text{ m.} \tag{56}$$

Therefore, from the relations (46) and (56) it results that the matter objects which at the end of radiation era were in the distance of gauge factor $a_{end}$ (54) and expanded at the velocity $v \sim c$ – under the hypothetical extensive assumption of mass increase of the homogeneous Universe in the result of the emergence of matter objects on the horizon (of the most remote visibility) – at present time would have to expand at the velocity (Skalský 1999)

$$v_{r(end-pres)} \sim 9 \times 10^6 \text{ ms}^{-1} \sim 3 \times 10^{-2} c. \tag{57}$$

Neither the Newton theory of gravitation, nor the Einstein general theory of relativity do not know the interaction, the force, or another reason, which could cause, or explain this hypothetical slowing down of the expansion of matter objects with the inertial-gravitational properties emerging on the Universe horizon (of the most remote visibility).

The gravitation is definitely insufficient and unsuitable for explanation of this hypothetical (and moreover homogeneous and isotropic) slowing down expansion of matter objects, constituting the matter component of the matter (mass)-space-time structure of the expansive and isotropic Universe.

In addition, according to the Einstein general theory of relativity, and according to the Newton theory of gravitation, the flat expansive universe in which the matter objects behave according to the relations (45) and (46), would not be able to begin its expansive evolution at all.

If we still admit – on the contrary to the above mentioned – the existence of such a flat expansive isotropic and homogeneous universe as a given fact, it would transform in the shortest possible time into the contractive one and next into the (final cosmological) singularity (Skalský 1997).

## 4. Conclusions

From the general geometrical point of view we can consider two principal possibilities of the permanent increase in mass of the expansive and isotropic Universe:

A. *Extensive*, as a result of emergence of matter objects on the Universe horizon of events.

B. *Intensive*, as a result of permanent origin (creation) of matter objects in "the visible Universe".

From above presented analysis of the physical results of the extensive assumption A, it results unambiguously that:

**The increase in mass of the observed expansive and isotropic relativistic Universe principally cannot be explained by the hypothetical extensive assumption on the emergence of matter objects on the hypothetical Newtonian-Euclidean horizon of events, because this assumption contradicts the theory of relativity, quantum mechanics, classical mechanics and observations** (Skalský 1997, 1999, 2000a, b, c, d, e)**.**

From the fact that geometrically only two principal possibilities can be considered, but one of them is physically not possible, it results unambiguously (apodictically) the necessity of the physical existence of the second possibility.

Therefore, from the presented facts it results unambiguously that:

**The increase in mass of the observed expansive and isotropic relativistic Universe can be realised only by the form of** *the hypothetical permanent constant maximum possible creation of matter*, determined by the relations *C* **(34) and** *D* **(35).**

**References**

Bondi, H. and Gold, T. 1948, *Month. Not. Roy. Astron. Soc.* **108**, 252.
Einstein, A. 1905, *Ann. Phys.* **17**, 891.
Friedmann, A. A. 1922, *Z. Phys.* **10**, 377.
Friedmann, A. A. 1924, *Z. Phys* **21**, 326.
Hawking, S. W. and Penrose, R. 1970, *Proc. Roy. Soc. London* **A314**, 529.
Hoyle, F. 1948, *Month. Not. Roy. Astron. Soc.* **108**, 372.
Hubble, E. P. 1929, *Proc. U.S. Nat. Acad. Sci.* **15**, 168.
Linde, A. D. 1990, *Particle Physics and Inflationary Cosmology*, Harwood Academic Publishers, Chur.
Monin, A. S., Polubarinova-Kochina, P. Ya. and Khlebnikov, V. I. 1989, *Cosmology, Hydrodynamics, Turbulence: A. A. Friedmann and Extension of His Scientific Heritage*, Nauka, Moscow.
Newton, I. 1687, *Philosophiae Naturalis Principia Mathematica*, Royal Society, London.
Planck, M. K. E. L. 1899, *Sitzber. Preuss. Akad. Wiss.* **26**, 440.
Robertson, H. P. 1935, *Ap. J.*, I, **82**, 284.
Robertson, H. P. 1936a, *Ap. J.* II, **83**, 187.
Robertson, H. P. 1936b, *Ap. J.* III, **83**, 257.
Skalský, V. 1991, *Astrophys. Space Sci.* **176**, 313. (Corrigendum: 1992, **187**, 163.)
Skalský, V. 1992, In: J. Dubnicka (ed.): *Philosophy, Natural Sciences and Evolution* (Proceedings of an Interdisciplinary Symposium, Smolenice, December 12-14, 1988), Slovak Academy of Sciences, Bratislava, p. 83.
Skalský, V. 1993, *Astrophys. Space Sci.* **201**, 3. (Corrigendum: 1994, **219**, 303.)
Skalský, V. 1994, *Astrophys. Space Sci.* **219**, 275.
Skalský, V. 1997, *DYNAMICS OF THE UNIVERSE in the Consistent and Distinguished Relativistic, Classically-Mechanical and Quantum-Mechanical Analyses*, Slovak Technical University, Bratislava.
Skalský, V. 1998, In: *ABSTRACTS* (JENAM-98, 7[th] European and 65[th] Czech Astronomical Society Meeting, Prague, September 9-12, 1998), European Space Agency, Noordwijk, p. 296.
Skalský, V. 1999, gr-qc/9912038
Skalský, V. 2000a, astro-ph/0001251
Skalský, V. 2000b, gr-qc/0001063
Skalský, V. 2000c, gr-qc/0002052
Skalský, V. 2000d, astro-ph/0001374
Skalský, V. 2000e, astro-ph/0003192
Walker, A. G. 1936, *Proc. Lond. Math. Soc.* **42**, 90.